\documentclass[11pt]{article}

\usepackage{amsmath}
\usepackage{amstext}
\usepackage{amsopn}
\usepackage{amsthm}
\usepackage{algorithm}
\usepackage[noend]{algorithmic}

\newcommand{\nb}[1]{\mathit{not}\,(#1)}
\newcommand{\Nb}[1]{\mathit{not}\bigl(#1\bigr)}
\newcommand{\nB}[1]{\mathit{not}\ #1}
\newcommand{\atleast}[1]{\mathit{Atleast}(#1)}
\newcommand{\atmost}[1]{\mathit{Atmost}(#1)}
\newcommand{\patoms}[1]{{#1}^+}
\newcommand{\natoms}[1]{{#1}^-}
\DeclareMathOperator{\Mid}{\big\vert}
\newcommand{\from}{\leftarrow}
\newcommand{\atoms}{\mathit{Atoms}}
\newcommand{\Atoms}[1]{\mathit{Atoms}(#1)}
\newcommand{\reduct}[2]{#1^{#2}}
\newcommand{\abs}[1]{\lvert#1\rvert}
\newcommand{\maps}{\rightarrow}

\theoremstyle{plain}
\newtheorem{theorem}{Theorem}
\newtheorem{lemma}[theorem]{Lemma}
\newtheorem*{lem}{Lemma}
\newtheorem{proposition}[theorem]{Proposition}

\theoremstyle{definition}
\newtheorem{definition}{Definition}
\theoremstyle{remark}
\newtheorem{example}{Example}

\title{Extending the Stable Model Semantics with More Expressive Rules}
\author{Patrik Simons%
\thanks{The financial support of the Academy of Finland and the
Helsinki Graduate School in Computer Science and Engineering is
gratefully acknowledged.} \\
Department of Computer Science and Engineering \\
Helsinki University of Technology, FIN-02015 HUT, Finland \\
\texttt{Patrik.Simons@hut.fi}, \texttt{http://www.tcs.hut.fi/\~{ }psimons}}
\date{}

\begin{document}

\maketitle

\begin{abstract}
The rules associated with propositional logic programs and the stable
model semantics are not expressive enough to let one write concise
programs. This problem is alleviated by introducing some new types of
propositional rules. Together with a decision procedure that has been
used as a base for an efficient implementation, the new rules supplant
the standard ones in practical applications of the stable model
semantics.
\end{abstract}

\section{Introduction}

Logic programming with the stable model semantics has emerged as a
viable method for solving constraint satisfaction
problems~\cite{MT:cs.LO/9809032,Niemela98:cnmr}. The state-of-the-art
system {\em smodels}~\cite{NS96:jicslp} can often handle non-stratified
programs with tens of thousands of rules. However, propositional logic
programs can not compactly encode several types of constraints. For
example, expressing the subsets of size $k$ of an $n$-sized set as
stable models requires on the order of $nk$ rules. In order to remedy
this problem, we improve upon the techniques of {\em smodels}, by
extending the semantics with some new types of propositional rules:
\begin{itemize}
\item choice rules for encoding subsets of a set,
\item constraint rules for enforcing cardinality limits on the
subsets, and
\item weight rules for writing inequalities over weighted linear sums.
\end{itemize}
The extended semantics is not based on subset-minimal models as is the
case for disjunctive logic programs. For instance, the choice rule is
more of a generalization of the disjunctive rule of the possible model
semantics~\cite{SI94}.

A system that computes the stable models of programs containing the
new rules has been implemented~\cite{smodels2}, and it has successfully
been applied to deadlock and reachability
problems in a class of Petri
nets~\cite{Heljanko:Tacas99}. Other problem domains,
such as planning and configuration, will benefit by the improved rules
as well. The system is based on {\em smodels} 1.10 from which it evolved.

The new rules and the stable model semantics are introduced in
Section~\ref{sec:stable}. A decision procedure for the extended syntax
is presented in Section~\ref{sec:decision}, and some important
implementation details are described in
Section~\ref{sec:implementation}. Experimental results are
found in Section~\ref{sec:experiments}. Readers not familiar with
monotonic functions should consult the appendix.

\section{The Stable Model Semantics}
\label{sec:stable}

Let $\atoms$ be a set of primitive propositions, or atoms, and consider
logic programs consisting of rules of the form
\[h \from a_1,\dotsc,a_n,\nB{b_1}, \dotsc ,\nB{b_m,}\]
where the head $h$ and the atoms $a_1,\dotsc,a_n,b_1,\dotsc,b_m$ in
the body are members of $\atoms$. Call the expression $\nB{b}$ a
not-atom --- atoms and not-atoms are referred to as literals.

The stable model semantics for a logic program $P$ is defined as
follows~\cite{GL88}. The reduct $\reduct{P}{A}$ of $P$ with respect
to the set of atoms $A$ is obtained by
\begin{enumerate}
\item deleting each rule in $P$ that has a not-atom $\nB{x}$ in its
body such that $x\in A$, and by
\item deleting all not-atoms in the remaining rules.
\end{enumerate}

\begin{definition}
A set of atoms $S$ is a stable
model of $P$ if and only if $S$ is the deductive closure of
$\reduct{P}{S}$ when the rules in $\reduct{P}{S}$ are seen as
inference rules.
\end{definition}

In order to facilitate the definition of more general forms of rules,
we introduce an equivalent characterization of the stable model
semantics.

\begin{proposition}
We say that $f_P : 2^\atoms\maps 2^\atoms$ is a closure if
\begin{multline*}
f_P(S) = \{ h \mid h \from
a_1,\dotsc,a_n,\nB{b_1},\dotsc,\nB{b_m} \in P, \\
 a_1,\dotsc,a_n\in f_P(S),\ b_1,\dotsc,b_m\not\in S \}.
\end{multline*}
Let
\[g_P(S) = \bigcap\{f_P(S) \mid \text{$f_P : 2^\atoms\maps 2^\atoms$
is a closure} \}.\]
Then, $S$ is a stable model of the program $P$ if and only if
\[S = g_P(S).\]
\end{proposition}

\begin{proof}
Note that the deductive closure of the reduct $\reduct{P}{S}$ is
a closure, and note that for every $f_P$ that is a closure, the
deductive closure of $\reduct{P}{S}$ is a subset of $f_P(S)$.
\end{proof}

A stable model is therefore a model that follows from itself by means
of the smallest possible closure. In other words, a stable model is a
supported model, and this is the essence of the semantics.

\begin{definition}
A basic rule $r$ is of the form
\[h \from a_1,\dotsc,a_n,\nB{b_1}, \dotsc ,\nB{b_m}\]
and is interpreted by the function
$f_r:2^\atoms\times 2^\atoms\maps 2^\atoms$
as follows.
\[f_r (S,C) = \{h \mid a_1,\dotsc,a_n\in C,\ b_1,\dotsc,b_m\not\in S\}.\]
\end{definition}
The function $f_r$ produces the result of a deductive step when
applied to a candidate stable model $S$ and its consequences $C$.
\begin{definition}
A constraint rule $r$ is of the form
\[h \from k\,\{ a_1,\dotsc,a_n, \nB{b_1}, \dotsc ,\nB{b_m}\}\]
and is interpreted by
\[
f_r (S,C) = \bigl\{h \Mid
\abs{\{a_1,\dotsc,a_n\}\cap C} +
\abs{\{b_1,\dotsc,b_m\} - S} \geq k\bigr\}.
\]
\end{definition}
The constraint rule can be used for testing the cardinality of a set
of atoms. The rule $h_1 \from 2\,\{ a,b,c,d \}$ states that $h_1$ is
true if at least $2$ atoms in the set $\{ a,b,c,d \}$ are true. The
rule $h_2 \from 1\,\{ \nB{a},\nB{b},\nB{c},\nB{d} \}$, on the other
hand, states that $h_2$ is true if at most $3$ atoms in the set are
true.
\begin{definition}
A choice rule $r$ is of the form
\[\{h_1,\dotsc,h_k\} \from a_1,\dotsc,a_n, \nB{b_1}, \dotsc ,\nB{b_m}\]
and is interpreted by
\[
f_r (S,C) = \bigl\{h \Mid h\in\{h_1,\dotsc,h_k\}\cap S,
a_1,\dotsc,a_n\in C,\ b_1,\dotsc,b_m\not\in S\bigr\}.
\]
\end{definition}
The choice rule is typically used when one wants to implement optional
choices. The rule $\{a\} \from b,\nB{c}$ declares that if $b$ is true
and $c$ is false, then $a$ is one or the other.
\begin{definition}
Finally, a weight rule $r$ is of the form
\[
h \from \{ a_1 = w_{a_1},\dotsc,a_n = w_{a_n},
\nB{b_1} = w_{b_1},\dotsc,\nB{b_m} = w_{b_m}\} \geq w,
\]
for $w_{a_i},w_{b_i} \geq 0$,
and is interpreted by
\[f_r (S,C) = \{h \mid \sum_{a_i\in C} w_{a_i} +
\sum_{b_i\not\in S} w_{b_i} \geq w\}.\]
\end{definition}
The weight rule is a generalization of the constraint rule. If every
literal in the body of a weight rule has weight $1$, then the rule
behaves precisely as a constraint rule.

\begin{definition}
Let $P$ be a set of rules. As before we say that
$f_P:2^\atoms\maps 2^\atoms$ is a closure if
\[f_P(S) = \bigcup_{r\in P} f_r\bigl(S,f_P(S)\bigr),\]
and we define
\[g_P(S) = \bigcap\{f_P(S) \mid \text{$f_P : 2^\atoms\maps 2^\atoms$
is a closure}\}.\]
Then, $S$ is a stable model of the program $P$ if and only if
\[S = g_P(S).\]
\end{definition}

The motivation for defining constraint, choice, and weight
rules is that they can be easily and efficiently implemented and that
they are quite expressive. For example, the constraint rule
\[h \from k\,\{ a_1,\dotsc,a_n,\nB{b_1},\dotsc,\nB{b_m}\}\]
replaces the program
\begin{multline*}
\{h\from a_{i_1},\dotsc,a_{i_{k_1}},
\nB{b_{j_1}},\dotsc,\nB{b_{j_{k_2}}} \mid k_1+k_2 = k, \\
1\leq i_1<\dotsb<i_{k_1}\leq n,\ 1\leq j_1<\dotsb<j_{k_2}\leq m\},
\end{multline*}
which contains $\binom{n+m}{k}$ rules.

Thus, a constraint rule guarantees that if the sum of the number of
atoms in its body that are in a stable model and the number of
not-atoms in its body that are not is at least $k$, then the head is
in the model. Similarly, if the body of a choice rule agrees with a
stable model, then the rule motivates the inclusion of any number of
atoms from its head. A weight rule
\[
h \from \{ a_1 = w_{a_1},\dotsc,a_n = w_{a_n},
\nB{b_1} = w_{b_1},\dotsc,\nB{b_m} = w_{b_m}\} \geq w,
\]
in turn, will force the head to be a member of a stable model $S$ if
\[\sum_{a_i\in S} w_{a_i}+\sum_{b_i\not\in S} w_{b_i} \geq w.\]

\begin{example}
The stable models of the program
\begin{align*}
\{a_1,\dotsc,a_n\} &\from \\
\mathit{false} &\from \{a_1 = w_1,\dotsc,a_n = w_n\} \geq w \\
\mathit{true} &\from \{a_1 = v_1,\dotsc,a_n = v_n\} \geq v
\end{align*}
containing the atom $\mathit{true}$ but not the atom $\mathit{false}$
correspond to the ways one can pack a subset of $a_1,\dotsc,a_n$ in a
bin such that the total weight is less than $w$ and the total value is
at least $v$. The individual weights and values of the items are given
by respectively $w_1,\dotsc,w_n$ and $v_1,\dotsc,v_n$.
\end{example}

\begin{example}
The satisfying assignments of the formula
\[(a\lor b\lor\neg c)\land (\neg a\lor b\lor\neg d)\land (\neg b\lor
c\lor d)\]
correspond to the stable models of the program
\begin{align*}
\{a,b,c,d\} &\from \\
\mathit{false} &\from \nB{a},\nB{b},c \\
\mathit{false} &\from a,\nB{b},d \\
\mathit{false} &\from b,\nB{c},\nB{d}
\end{align*}
that do not contain $\mathit{false}$.
\end{example}

\section{The Decision Procedure}
\label{sec:decision}

For an atom $a$, let $\nb{a} = \nB{a}$, and for a not-atom $\nB{a}$,
let \[\nb{\nB{a}} = a.\] For a set of literals $A$, define
\[\nb{A} = \{ \nb{a} \mid a\in A\}.\]
Let $\patoms{A} = \{a\in\atoms\mid a\in A\}$ and let
$\natoms{A} = \{a\in\atoms\mid \nB{a}\in A\}$. Define
$\Atoms{A} = \patoms{A}\cup\natoms{A}$, and for a program $P$, define
$\Atoms{P} = \Atoms{L}$, where $L$ is the set of literals that appear
in the program.

A set of literals $A$ is said to cover a set of atoms $B$ if
$B\subseteq\Atoms{A}$, and $B$ is said to agree with $A$ if
\[\patoms{A}\subseteq B \quad\text{and}\quad
\natoms{A}\subseteq\atoms-B.\]

Algorithm~\ref{algorithm:smodels} displays a decision procedure for
the stable model semantics. The function $\mathit{smodels}(P,A)$
returns true whenever there is a stable model of $P$ agreeing with
$A$, and it relies on the three functions $\mathit{expand}(P,A)$,
$\mathit{conflict}(P,A)$, and $\mathit{lookahead}(P,A)$.

\begin{algorithm}
\caption{A decision procedure for the stable model semantics}
\begin{algorithmic}\label{algorithm:smodels}
\medskip
\item[\textbf{function} $\mathit{smodels}(P,A)$]
 \STATE $A':=\mathit{expand}(P,A)$
 \IF{$\mathit{conflict}(P,A')$}
  \STATE return false
 \ELSIF{$A'$ covers $\Atoms{P}$}
  \STATE return true \COMMENT{$\patoms{A'}$ is a stable model}
 \ELSE
  \STATE $x:=\mathit{lookahead}(P,A')$
  \IF{$\mathit{smodels}(P,A'\cup \{x\})$}
   \STATE return true
  \ELSE
   \STATE return $\mathit{smodels}\bigl(P,A'\cup\{\nb{x}\}\bigr)$.
  \ENDIF
 \ENDIF
\medskip
\item[\textbf{function} $\mathit{expand}(P,A)$]
 \REPEAT
  \STATE $A':=A$
  \STATE $A:=\atleast{P,A}$
  \STATE $A:=A\cup\{\nB{x}\mid x\in\Atoms{P}$ and
$x\not\in\atmost{P,A}\}$
 \UNTIL{$A = A'$}
 \STATE return $A$.
\medskip
\item[\textbf{function} $\mathit{conflict}(P,A)$]
 \STATE \COMMENT{Precondition: $A = \mathit{expand}(P,A)$}
 \IF{$\patoms{A}\cap\natoms{A} \neq \emptyset$}
  \STATE return true
 \ELSE
  \STATE return false.
 \ENDIF
\medskip
\item[\textbf{function} $\mathit{lookahead}(P,A)$]
 \STATE $B := \Atoms{P}-\Atoms{A}$; $B := B\cup\nb{B}$
 \WHILE{$B \neq \emptyset$}
  \STATE Take any literal $x\in B$
  \STATE $A' := \mathit{expand}(P,A\cup\{x\})$
  \IF{$\mathit{conflict}(P,A')$}
   \STATE return $x$
  \ELSE
   \STATE $B:=B-A'$
  \ENDIF
 \ENDWHILE
 \STATE return $\mathit{heuristic}(P,A)$.
\medskip
\end{algorithmic}
\end{algorithm}

Let $A' = \mathit{expand}(P,A)$. We assume that
\begin{description}
\item[E1] $A\subseteq A'$ and that
\item[E2] every stable model of $P$ that agrees with $A$
also agrees with $A'$.
\end{description}
Moreover, we assume that the function $\mathit{conflict}(P,A)$
satisfies the two conditions
\begin{description}
\item[C1] if $A$ covers $\Atoms{P}$ and there is no stable model that
agrees with $A$, then $\mathit{conflict}(P,A)$ returns true, and
\item[C2] if $\mathit{conflict}(P,A)$ returns true, then there is no
stable model of $P$ that agrees with $A$.
\end{description}
In addition, $\mathit{lookahead}(P,A)$ is expected to return literals
not covered by $A$.

\begin{theorem}
Let $P$ be a set of rules and let $A$ be a set of literals. Then,
there is a stable model of $P$ agreeing with $A$ if and only if
$\mathit{smodels}(P,A)$ returns true.
\end{theorem}

\begin{proof}
Let $\mathit{nc}(P,A) = \Atoms{P}-\Atoms{A}$ be the atoms not covered
by $A$. We prove the claim by induction on the size of
$\mathit{nc}(P,A)$.

Assume that the set $\mathit{nc}(P,A) = \emptyset$. Then, $A'$ covers
$\Atoms{P}$ by E1 and $\mathit{smodels}(P,A)$ returns true if and only
if $\mathit{conflict}(P,A')$ return false. By E2, C1, and C2, this
happens precisely when there is a stable model of $P$ agreeing with $A$.

Assume $\mathit{nc}(P,A)\neq\emptyset$. If $\mathit{conflict}(P,A')$
returns true, then $\mathit{smodels}(P,A)$ returns false and by E2 and 
C2 there is no stable model agreeing with $A$. On the other hand, if
$\mathit{conflict}(P,A')$ returns false and $A'$ covers $\Atoms{P}$,
then $\mathit{smodels}(P,A)$ returns true and by E2 and C1 there is a
stable model that agrees with $A$. Otherwise, induction together with
E1 and E2 show that $\mathit{smodels}(P,A'\cup\{x\})$ or
$\mathit{smodels}\bigl(P,A'\cup\{\nb{x}\}\bigr)$ returns true if and
only if there is a stable model agreeing with $A$.
\end{proof}

Let $S$ be a stable model of $P$ agreeing with the set of literals $A$.
Then, $f_r(S,S)\subseteq S$ for $r\in P$, and we make the following
observations. Let
\[\mathit{min}_r(A) = \negthickspace
\bigcap_{\substack{\patoms{A}\subseteq C \\
\natoms{A}\cap C=\emptyset}} \negthickspace f_r(C,C)\]
be the inevitable consequences of $A$, and let
\[\mathit{max}_r(A) = \negthickspace
\bigcup_{\substack{\patoms{A}\subseteq C \\
\natoms{A}\cap C=\emptyset}} \negthickspace f_r(C,C)\]
be the possible consequences of $A$.
Then,
\begin{enumerate}
\item for all $r\in P$, $S$ agrees with $\mathit{min}_r(A)$,
\item if there is an atom $a$ such that for all $r\in P$,
$a\not\in\mathit{max}_r(A)$, then $S$ agrees with $\{\nB{a}\}$,
% The following is monotonic as long as A\subseteq S
\item if the atom $a\in A$, if there is only one $r\in P$ for which
$a\in\mathit{max}_r(A)$, and if there exists a literal $x$ such that
$a\not\in\mathit{max}_r(A\cup\{x\})$, then $S$ agrees with
$\{\nb{x}\}$, and
\item if $\nB{a}\in A$ and if there exists a literal $x$ such that
for some $r\in P$, $a\in\mathit{min}_r(A\cup\{x\})$, then $S$ agrees
with $\{\nb{x}\}$.
\end{enumerate}
The four statements help us deduce additional literals that are in
agreement with $S$. Define $\atleast{P,A}$ as the smallest set of literals
containing $A$ that can not be enlarged using 1--4 above, i.e., let
$\atleast{P,A}$ be the least fixed point of the operator
\[\begin{align*}
f(B) = A&\cup B\cup\{a\in \mathit{min}_r(B) \mid r\in P\} \\
&\cup\{\nB{a}\mid \text{$a\in\Atoms{P}$ and for all $r\in P$,
$a\not\in\mathit{max}_r(B)$}\} \\
&\cup\bigl\{\nb{x}\Mid \text{there exists $a\in B$ such that
$a\in\mathit{max}_r(B)$} \\
&\qquad\qquad\qquad\text{for only one $r\in P$ and
$a\not\in\mathit{max}_r(B\cup\{x\})$}\bigr\} \\
&\cup\bigl\{\nb{x}\Mid \text{there exists $\nB{a}\in B$ and $r\in P$
such that} \\
&\qquad\qquad\qquad a\in\mathit{min}_r(B\cup\{x\})\bigr\}.
\end{align*}\]

\begin{lemma}\label{lemma:atleast}
The function $\atleast{P,A}$ is monotonic in its second argument.
\end{lemma}

\begin{proof}
Observe that the function $\mathit{min}_r(B)$ is monotonic and that
the function $\mathit{max}_r(B)$ is anti-monotonic. Hence, 
\begin{gather*}
\{a\in \mathit{min}_r(B) \mid r\in P\}, \\
\{\nB{a}\mid \text{$a\in\Atoms{P}$ and for all $r\in P$,
$a\not\in\mathit{max}_r(B)$}\}, \\
\intertext{and}
\bigl\{\nb{x}\Mid \text{there exists $\nB{a}\in B$ and $r\in P$
such that $a\in\mathit{min}_r(B\cup\{x\})$}\bigr\}
\end{gather*}
are monotonic with respect to $B$. Assume that there exists $a\in B$
such that $a\in\mathit{max}_r(B)$ for only one $r\in P$ and
$a\not\in\mathit{max}_r(B\cup\{x\})$. If $B\subseteq B'$ and 
$a\not\in\mathit{max}_r(B')$, then
\[\nB{a}\in\{\nB{a}\mid \text{$a\in\Atoms{P}$ and for all $r\in P$,
$a\not\in\mathit{max}_r(B')$}\} \subseteq f(B').\]
Consequently, both $a,\nB{a}\in f(B')$ and therefore
\begin{gather*}
\mathit{min}_r\bigl(f(B')\bigr) = \atoms \\
\intertext{and}
\mathit{max}_r\bigl(f(B')\bigr) = \emptyset.
\end{gather*}
It follows that
$f\bigl(f(B')\bigr) = \Atoms{P}\cup\Nb{\Atoms{P}}$. Thus, $f^2$
is monotonic and has a least fixed point. Finally, notice that $f$ has 
the same fixed points as $f^2$.
\end{proof}

We conclude,

\begin{proposition}
If the stable model $S$ of $P$ agrees with $A$, then $S$ agrees with
$\atleast{P,A}$.
\end{proposition}

Furthermore, we can bound the stable models from above.

\begin{proposition}
For a choice rule $r$ of the form
\[\{h_1,\dotsc,h_k\} \from a_1,\dotsc,a_n, \nB{b_1}, \dotsc ,\nB{b_m},\]
let
\[f_r' (S,C) = \bigl\{h\in\{h_1,\dotsc,h_k\} \Mid
a_1,\dotsc,a_n\in C,\ b_1,\dotsc,b_m\not\in S\bigr\},\]
and for any other type of rule, let $f_r' (S,C) = f_r (S,C)$.
Let $S$ be a stable model of $P$ that agrees with $A$. Define
$\atmost{P,A}$ as the least fixed point of
\[f'(B) = \bigcup_{r\in P} f_r'(\patoms{A},B-\natoms{A})-\natoms{A}.\]
Then, $S\subseteq \atmost{P,A}$.
\end{proposition}

\begin{proof}
Note that $f_r'(S,C)$ is anti-monotonic in its first argument, i.e.,
$S\subseteq S'$ implies $f_r'(S',C)\subseteq f_r'(S,C)$, and monotonic
in its second argument. Fix a program $P$, a stable model $S$ of $P$,
and a set of literals $A$ such that $S$ agrees with $A$. Define
\[f(B) = \bigcup_{r\in P} f_r(S,B)\]
and
\[f'(B) = \bigcup_{r\in P} f_r'(\patoms{A},B-\natoms{A})-\natoms{A}.\]
Let $L$ be the least fixed point of $f'$. Since $S$ agrees with $A$,
\[f_r(S,S\cap L)\subseteq f_r'(\patoms{A},S\cap L -
\natoms{A})-\natoms{A},\]
and $f(S\cap L)\subseteq f'(S\cap L) \subseteq L$. Hence, the least
fixed point of $f(S\cap\cdot)$, which is equal to the least fixed
point of $f$, is a subset of $L$. In other words, $S\subseteq L$.
\end{proof}

It follows that $\mathit{expand}(P,A)$ satisfies the conditions E1 and 
E2. The function $\mathit{conflict}(P,A)$ obviously fulfills C2, and 
the next proposition shows that also C1 holds.

\begin{proposition}
If $A = \mathit{expand}(P,A)$ covers the set $\Atoms{P}$ and
$\patoms{A}\cap\natoms{A} = \emptyset$, then $\patoms{A}$ is a stable
model of $P$. 
\end{proposition}

\begin{proof}
Assume that $A = \mathit{expand}(P,A)$ covers $\Atoms{P}$ and that
$\patoms{A}\cap\natoms{A} = \emptyset$. Then,
$\patoms{A} = \atmost{P,A}$. As
$f_r(\patoms{A},B)\subseteq\mathit{min}_r(A)\subseteq A$ for
$B\subseteq \patoms{A}$,
\[f_r(\patoms{A},B) = f_r'(\patoms{A},B-\natoms{A})-\natoms{A}\]
for every $B\subseteq\patoms{A}$. Thus, $\patoms{A}$ is the least
fixed point of \[f(B) = \bigcup_{r\in P} f_r(\patoms{A},B),\] from
which we infer that $\patoms{A}$ is a stable model of $P$.
\end{proof}

\subsection{Looking Ahead and the Heuristic}

Besides $\atleast{P,A}$ and $\atmost{P,A}$, there is a third way to
prune the search space. If the stable model $S$ agrees with $A$ but
not with $A\cup\{x\}$ for some literal $x$, then $S$ agrees with
$A\cup\{\nb{x}\}$. One can therefore avoid futile choices if one looks
ahead and tests whether $A\cup\{x\}$ gives rise to a conflict for some
literal $x$. Since $x'\in\mathit{expand}(P,A\cup\{x\})$ implies
\[\mathit{expand}(P,A\cup\{x'\})\subseteq
\mathit{expand}(P,A\cup\{x\})\]
due to the monotonicity of $\atleast{P,A}$ and $\atmost{P,A}$, it is
not even necessary to examine all literals not covered by $A$. That
is, if we have tested $x$, then we do not have to test the literals in 
$\mathit{expand}(P,A\cup\{x\})$.

When looking ahead fails to find a literal that causes a conflict, one
falls back on a heuristic. For a literal $x$, let
\[A_p = \mathit{expand}(P,A\cup\{x\})\]
and
\[A_n = \mathit{expand}\bigl(P,A\cup\{\nb{x}\}\bigr).\]
Assume that the search space is a full binary tree of height $H$,
and let $p = \abs{A_p-A}$ and $n = \abs{A_n-A}$. Then,
\[2^{H-p}+2^{H-n} = 2^H\frac{2^n+2^p}{2^{p+n}}\]
is an upper bound on the size of the remaining search space.
Minimizing this number is equal to minimizing
\[\log \frac{2^n+2^p}{2^{p+n}} = \log(2^n+2^p) - (p+n).\]
Since
\[2^{\max(n,p)} < 2^n+2^p \leq 2^{\max(n,p)+1}\]
is equivalent to
\[\max(n,p) < \log(2^n+2^p) \leq \max(n,p)+1\]
and
\[-\min(n,p) < \log(2^n+2^p) - (p+n)\leq 1-\min(n,p),\]
it suffices to maximize $\min(n,p)$. If two different literals have
equal minimums, then one chooses the one with the greater maximum,
$\max(n,p)$.

\section{Implementation Details}
\label{sec:implementation}

The deductive closures $\atleast{P,A}$ and $\atmost{P,A}$ can both
be implemented using two versions of a linear time algorithm of
Dowling and Gallier~\cite{DG84}. The basic algorithm associates with each
rule a counter that keeps track of how many literals in the body of
a rule are not included in a partially computed closure. If a counter
reaches zero, then the head of the corresponding rule is included in
the closure. From the inclusion follows changes in other counters, and
in this manner is membership in the closure propagated.

We begin with basic rules of the form
\[h \from a_1,\dotsc,a_n,\nB{b_1}, \dotsc ,\nB{b_m}.\]
For every rule $r$ we create a literal counter $r.\mathit{literal}$,
which is used as above, and an inactivity counter
$r.\mathit{inactive}$. If the set $A$ is a partial closure, then the
inactivity counter records the number of literals in the body of $r$
that are in $\nb{A}$. The counter $r.\mathit{inactive}$ is therefore
positive, and the rule $r$ is inactive, if one can not now nor later
use $r$ to deduce its head. For every atom $a$ we create a head
counter $a.\mathit{head}$ that holds the number of active rules with
head $a$.

Recall that a literal can be brought into $\atleast{P,A}$ in four
different ways. We handle the four cases with the help of the three
counters.
\begin{enumerate}
\item If $r.\mathit{literal}$ reaches zero, then the head of $r$ is
added to the closure.
\item If $a.\mathit{head}$ reaches zero, then $\nB{a}$ is added to the
closure.
\item If $a.\mathit{head}$ is equal to one and $a$ is in the closure,
then every literal in the body of the only active rule with head $a$
is added to the closure. 
\item Finally, if $a$ is the head of $r$, if $\nB{a}$ is in the
closure, and if $r.\mathit{literal} = 1$ and
$r.\mathit{inactive} = 0$, then there is precisely one literal $x$ in
the body of $r$ that is not in the closure, and $\nb{x}$ is added to
the closure.
\end{enumerate}

Constraint rules and choice rules are easily incorporated into the
same framework. Specifically, one does neither use the first nor the
fourth case together with choice rules, and one does not compare the
literal and inactivity counters of a constraint rule
$h \from k\,\{ a_1,\dotsc,a_n, \nB{b_1}, \dotsc ,\nB{b_m}\}$
with zero but with $m+n-k$. A weight rule
\[h \from \{ a_1 = w_{a_1},\dotsc,a_n = w_{a_n},
\nB{b_1} = w_{b_1},\dotsc,\nB{b_m} = w_{b_m}\} \geq w,\]
is managed using the upper and lower bound of the sum of the weights
in its body. Given a set of literals $A$, the lower bound is
\[\sum_{a_i\in \patoms{A}}w_{a_i} + \sum_{b_i\in \natoms{A}}w_{b_i}\]
and the upper bound is
\[\sum_{a_i\not\in\natoms{A}}w_{a_i} + \sum_{b_i\not\in
\patoms{A}}w_{b_i}.\]
If the upper bound is less than $w$, then the rule is inactive, and if
the lower bound is at least $w$, then the head is in the closure.

Notice that the implementation provides for incremental updates to
the closure $\atleast{P,A}$ as $A$ changes. This is crucial
for achieving a high performance.

Since the function $\atmost{P,A}$ is anti-monotonic, it will shrink as
$A$ grows. It is no good computing $\atmost{P,A}$ anew each time $A$
is modified. Instead all atoms that might not be in the newer and
smaller closure are found using a variant of the basic algorithm. By 
inspecting these atoms it is possible to decide which ones must be in
the closure, and then the basic algorithm can again be used to
compute the final closure. A small example will make the method
clear.

\begin{example}
Suppose $P$ is the program
\begin{align*}
a &\from b & a &\from \nB{c} \\
b &\from a & a &\from \nB{d},
\end{align*}
and suppose $A$ has changed from the empty set to $\{d\}$. Then, we
have already computed $\atmost{P,\emptyset} = \{a,b\}$, and we want to
find $\atmost{P,A}$. If $r$ is the rule $a \from \nB{d}$, then the
counter of $r$ is at first zero and then changes to one as $d$ becomes
a member of $A$. Therefore, we deduce that $a$ is possibly not a part
of the new closure. The basic algorithm proceeds to increment the
counters of $b \from a$, removing $b$, and $a \from b$, where it
stops. At this point the counter of the rule $a \from \nB{c}$ is
still zero, and we note that $a$ must be part of the
closure. Including $a$ causes the counter of $b \from a$ to
decrease to zero. Consequently, $b$ is added to the closure and the
counter of $a \from b$ is decremented. Since nothing more remains to
be done, the final closure is $\{a,b\}$.
\end{example}

One can argue, in this particular example, that $a$ follows from the
rule $a \from \nB{c}$ and need not be removed in the first stage of
the procedure. However, in general it is not possible to decide
whether an atom is in the final closure by inspecting the rules of
which it is a head. Notwithstanding, we can make improvements based
upon this observation.

For every atom $a$, create a source pointer whose mission is to point
to the first rule that causes $a$ to be included in the
closure. During the portion of the computation when atoms are removed
from the closure, we only remove atoms which are to be removed due to
a rule in a source pointer. For if the rule in a source pointer
does not justify the removal of an atom, then the atom is reentered
into the closure in the second phase of the computation. In practice,
this simple trick yields a substantial speedup of the computation of
$\atmost{P,A}$.

\section{Experiments}
\label{sec:experiments}

We will search for sets of binary words of length $n$ such
that the Hamming distance between any two words is at least $d$.
The size of the largest of these sets is denoted by $A(n,d)$. For
example, $A(5,3) = 4$ and any $5$-bit one-error-correcting code
contains at most 4 words. One such code is
$\{00000,00111,11001,11110\}=\{0,7,25,30\}$. Finding codes becomes
very quickly very hard. For instance, it was only recently proved that 
$A(10,3)=72$~\cite{OBK99}.

Construct a program that includes a rule
\[w_i \from \nB{w_{j_1}},\dotsc,\nB{w_{j_k}}\]
for every word $i=0,\dotsc,2^n$ such that $j_1,\dotsc,j_k$ are the
words whose distance to $i$ is positive and less than $d$. Then, the
stable models of the program are the maximal codes with Hamming
distance $d$. Add the rule
\[\mathit{true} \from m\,\{w_0,\dotsc,w_{2^n}\}\]
and every model containing $\mathit{true}$ is a code of size at least
$m$. For the purpose of making the problem a bit more tractable, we
only consider codes that include the zero word.

The test results are tabulated below. The minimum, maximum, and
average times are given in seconds and are calculated from ten runs on
randomly shuffled instances of the program. All tests where run under
Linux~2.2.6 on a 233MHz Pentium~II with 128MB of memory.

\begin{center}
\small
\begin{tabular}{|@{\hspace{2pt}}l@{\hspace{2pt}}|@{\hspace{2pt}}r@{\hspace{2pt}}|@{\hspace{2pt}}r@{\hspace{2pt}}|@{\hspace{2pt}}r@{\hspace{2pt}}|}
\hline
Problem & Min & Max & Average \\
\hline
$A(5,3)\geq 4$ & 0.01 & 0.02 & 0.02 \\
$A(5,3)< 5$ & 0.00 & 0.02 & 0.02 \\
$A(6,3)\geq 8$ & 0.02 & 0.04 & 0.03 \\
$A(6,3)< 9$ & 0.16 & 0.18 & 0.17 \\
$A(7,3)\geq 16$ & 0.14 & 14.19 & 6.77 \\
$A(7,3)< 17$ & 69.08 & 72.29 & 70.55 \\
$A(8,3)\geq 20$ & 6.39 & 202.41 & 55.98 \\ \cline{2-4}
$A(8,3)< 21$ & \multicolumn{3}{c|}{$>1$ week} \\
\hline
\end{tabular}
\quad
\begin{tabular}{|@{\hspace{2pt}}l@{\hspace{2pt}}|@{\hspace{2pt}}r@{\hspace{2pt}}|@{\hspace{2pt}}r@{\hspace{2pt}}|@{\hspace{2pt}}r@{\hspace{2pt}}|}
\hline
Problem & Min & Max & Average \\
\hline
$A(6,5)\geq 2$ & 0.02 & 0.03 & 0.03 \\
$A(6,5)< 3$ & 0.02 & 0.03 & 0.02 \\
$A(7,5)\geq 2$ & 0.05 & 0.07 & 0.06 \\
$A(7,5)< 3$ & 0.04 & 0.07 & 0.06 \\
$A(8,5)\geq 4$ & 0.29 & 0.36 & 0.34 \\
$A(8,5)< 5$ & 2.64 & 2.75 & 2.71 \\
$A(9,5)\geq 6$ & 3.18 & 8.71 & 4.81 \\
$A(9,5)< 7$ & 1127.03 & 1162.10 & 1145.85 \\
\hline
\end{tabular}
\end{center}

\section{Conclusion}
\label{sec:conclusion}

We have presented some new and more expressive propositional rules for
the stable model semantics. A decision procedure, which has been used
as a base for an efficient implementation, has also been described.
We note that the decision problem for the extended semantics is
$\mathit{NP}$-complete, as a proposed stable model can be tested in
polynomial time. Accordingly, the exponential worst case
time-complexity of the decision procedure comes as no surprise.

The literals that $\mathit{smodels}(P,A)$ can branch on are, in this
paper, the literals that do not cover $\Atoms{P}-\Atoms{A}$. In
previous work, for instance in Niemel\"a and
Simons~\cite{NS96:jicslp,S97}, the eligible literals have also been
required to appear in the form of not-atoms in the program. This
additional restriction can reduce the search space, and a similar
requirement is, of course, also possible here. The question of which
literals one necessarily must consider as branch points is left to
future research.

\section*{Appendix}

Let $X$ be a set and let $f:2^X\maps 2^X$ be a function. If
$A\subseteq B$ implies $f(A)\subseteq f(B)$, then $f$ is monotonic.

\newcommand{\lfp}[1]{\mathit{lfp}\left(#1\right)}
\newcommand{\gfp}[1]{\mathit{gfp}\left(#1\right)}

\begin{lem}
Let $f:2^X\maps 2^X$ be a monotonic function, and let
$A\subseteq X$. If $f(A) \subseteq A$, then $\lfp{f}\subseteq A$,
where $\lfp{f}$ denotes the least fixed point of $f$.
\end{lem}

\begin{proof}
Define
\[S = \bigcap_{f(A)\subseteq A} A \qquad\qquad
\bigl(f(X)\subseteq X\bigr).\]
Then, $f(A)\subseteq A$ implies $S\subseteq A$, which in turn implies
$f(S)\subseteq f(A)$ by the monotonicity of $f$. Hence,
$f(S)\subseteq A$, and consequently
\[f(S) = \bigcap_{f(A)\subseteq A} f(S) \subseteq
\bigcap_{f(A)\subseteq A} A = S.\]

Now, $f(S)\subseteq S$ implies $f\bigl(f(S)\bigr)\subseteq f(S)$,
which by the definition of $S$ implies $S\subseteq f(S)$. Thus,
$S = f(S)$. Moreover, for any fixed point $A$,
\[f(A) \subseteq A \quad\text{implies}\quad S\subseteq A,\] 
and hence $\lfp{f} = S$ by definition.
\end{proof}

Similarly, $A\subseteq f(A)$ implies $A\subseteq \gfp{f}$ for the
greatest fixed point of $f$. Notice that if $X$ is finite, then
$\lfp{f} = f^n(\emptyset)$ for some $n\leq\abs{X}$ since
$f(\emptyset)\subseteq\lfp{f}$. Furthermore, observe that if we are
given $k$ monotonic functions $f_1,\dotsc,f_k$, then the least fixed
point of
\[g(A) = \bigcup_{i=1}^k f_i(A)\]
is the limit of any nest
\[A_{n+1} = A_n\cup f_{i(n)}(A_n),\quad \text{$A_0=\emptyset$ and
$f_{i(n)}(A_n)\subseteq A_n \Rightarrow \forall j f_j(A_n)\subseteq A_n$.}\]
In other words, the least fixed point of $g$ can be computed by
repeated applications of $f_1,\dotsc,f_k$.

\end{document}